\documentclass[letterpaper, 10pt, conference]{ieeeconf}
\IEEEoverridecommandlockouts
\usepackage{graphics,amsmath,amssymb,xcolor,mdframed,cite,booktabs,multirow}
\usepackage{pgfplots}
\usepackage[caption=false]{subfig}

\pgfmathdeclarefunction{gauss}{2}{%
  \pgfmathparse{1/(#2*sqrt(2*pi))*exp(-((x-#1)^2)/(2*#2^2))}%
}

\newtheorem{thm}{Theorem}
\newtheorem{defn}{Definition}
\newtheorem{lem}{Lemma}
\newtheorem{cor}{Corollary}
\newtheorem{ex}{Example}
\newenvironment{theorem}{\medskip\begin{thm}}{\end{thm}\medskip}

\newenvironment{lemma}{\medskip\begin{lem}}{\end{lem}\medskip}
\newenvironment{corollary}{\medskip\begin{cor}}{\end{cor}\medskip}

\newcommand{\logn}{\ln\left(\frac{N}{\eps_1}\right)}
\newcommand{\logk}{\ln\left(\frac{K}{\eps_2}\right)}

\newcommand{\etan}{\eta_{\text{non}}}
\newcommand{\etaa}{\eta_{\text{ada}}}

\newcommand{\var}{\operatorname{Var}}
\newcommand{\expn}{\mathbb{E}}

\newcommand{\eps}{\varepsilon}
\newcommand{\act}{\mathcal{A}}

\title{\LARGE \bf Efficient Probabilistic Group Testing Based on Traitor Tracing}

\author{Thijs Laarhoven$^{1}$% <-this % stops a space
\thanks{$^{1}$Thijs Laarhoven is with the Department of Mathematics and Computer Science, Eindhoven University of Technology, The Netherlands.\protect\\% 
E-mail: {\tt\small mail@thijs.com}.}%
}

\begin{document}

\maketitle
\thispagestyle{empty}
\pagestyle{empty}

%++++++++++++++++++++++++++++++++++++++++++++++++++++++++++++++++++++++++++++++%

\begin{abstract}
Inspired by recent results from collusion-resistant traitor tracing, we provide a framework for constructing efficient probabilistic group testing schemes. In the traditional group testing model, our scheme asymptotically requires $T \sim 2 K \ln N$ tests to find (with high probability) the correct set of $K$ defectives out of $N$ items. The framework is also applied to several noisy group testing and threshold group testing models, often leading to improvements over previously known results, but we emphasize that this framework can be applied to other variants of the classical model as well, both in adaptive and in non-adaptive settings. 
\end{abstract}

%++++++++++++++++++++++++++++++++++++++++++++++++++++++++++++++++++++++++++++++%

\section{INTRODUCTION}

%..............................................................................%

\subsection{Group testing}

Suppose a large population has to be tested for a certain illness, to determine which people are ill. One way to do this is to take blood samples from each person and test these samples one by one. However, if only few people are ill, many tests are wasted on non-infected people. It may then be advantageous to test bigger pools of several blood samples with \textit{group tests}. If one of the tested people in a pool is ill, the test will come back positive and further tests are required, but if the test comes back negative, we may conclude that none of the people in the tested group are ill and many tests are saved. \textit{Group testing} concerns the identification of a small subset of $K$ defectives hidden among $N$ total items, using the aforementioned group tests. The goal of group testing is to minimize the number of group tests $T$ required to identify the defectives, by carefully choosing which groups to test. 

\paragraph{Adaptive group testing}
In 1943, Dorfman~\cite{dorfman43} published a seminal paper studying practical ways of testing many blood samples of soldiers for syphilis, which is widely regarded as the first work on group testing. In the decades to follow, a lot of research was done in the area of \textit{adaptive} group testing, where many sequential rounds of testing are considered, and the selection of samples for the next pool may be influenced by the results of previous group tests. In this adaptive setting, using a binary search, $T = K \lceil \log_2 N \rceil$ tests suffice to detect $K$ defectives in a sample of size $N$. Up to a constant factor, this number of tests is optimal.

\paragraph{Non-adaptive group testing} 
For practical and economical reasons, the focus of later work in group testing shifted more towards the \textit{non-adaptive} setting, where many tests are run in parallel in one or few rounds. With certain combinatorial designs it is possible to find all $K$ defectives in one round with $T = O(K^2 \log (N/K))$ tests~\cite{dyachkov82}, while a lower bound of $T = \Omega(K^2 \log N/\log K)$~\cite{dyachkov89} shows that this number of tests is nearly optimal, when one round of tests is done and when the group testing algorithm always has to identify the correct subset of defectives. If we allow for a small positive probability $\eps$ of not detecting the right set of defectives, then even in one round of tests, $T = O(K \log N)$ parallel tests suffice to isolate all defectives with high probability. Together with the lower bound of $T \geq K \log_2 N$ for large $N$~\cite{sebo85}, this shows that $T = \Theta(K \log N)$ is optimal. Chan et al.~\cite{chan11} recently gave a computationally efficient algorithm that belongs in the latter category that uses $T = e K \log(N/\eps)$ tests to get a success probability of at least $1 - \eps$.

\paragraph{Variants}
Besides the pure group testing model, variants have also been studied, such as noisy group testing~\cite{atia09, atia12, chan11, cheraghchi09, cheraghchi11, sejdinovic10} and threshold group testing~\cite{chan13, damaschke06, lebedev10}. In these models a positive outcome of a test is not equivalent to at least one defective being present in the tested group, as there may be a small probability of making a mistake in the testing procedure, or because the test might not come back positive if very few defectives are present in the tested group. The trivial adaptive group testing algorithm of doing a binary search does not work in these models, and so even finding an efficient adaptive group testing scheme in these models is not easy.

%..............................................................................%

\subsection{Collusion-resistant traitor tracing}

A completely different, but in fact closely related area of research is that of collusion-resistant traitor tracing. To protect digital content from unauthorized redistribution, copyright holders embed watermarks in the content such that, if an illegal copy is made and distributed, the watermark can be linked to the responsible user. Things become more complicated when several pirates collude, and start mixing their copies to create a new pirated copy of the content that does not match any of their copies of the content exactly. If in some segment of the data all pirates receive the same watermarked version, the \textit{marking assumption}~\cite{boneh98} says that they are forced to output this version of the content. However, if they receive several different versions, they may choose any of them to output. \textit{Traitor tracing} concerns assigning watermarks to $N$ users in such a way that, even if $K$ users mix their copies as described above, we may still be able to find the colluders. The goal of traitor tracing is to minimize the number of segments $T$ needed to trace (part of) the coalition, by carefully choosing which watermarked versions of each segment to assign to each user.

\setcounter{paragraph}{0} 
\paragraph{Static (non-adaptive) traitor tracing}
Work on traitor tracing started only in the late $20$th century. In many of the early constructions, the number of segments required was polynomial in $N$, until Boneh and Shaw~\cite{boneh98} gave an efficient construction that uses $T = O(K^4 \log(N/\eps))$ segments to find at least one of the colluders with high probability in the static (non-adaptive) setting. Upper and lower bounds on $T$ were further improved until in 2003, Tardos~\cite{tardos03} showed that $T = O(K^2 \log(N/\eps))$ segments are both necessary and sufficient. In the same paper he presented an efficient scheme that achieves this lower bound up to a constant factor. Later research focused on establishing the exact (asymptotic) lower bound~\cite{huang12}, which turned out to be $T \gtrsim 2 K^2 \ln N$, and decreasing the upper bounds by improving Tardos' scheme~\cite{blayer08, skoric08, laarhoven12dcc, oosterwijk13}, which eventually lead to an asymptotic bound of $T \sim \frac{1}{2} \pi^2 K^2 \ln(N/\eps)$. 

\paragraph{Dynamic (adaptive) traitor tracing}
While the above results are based on the static setting where the assignment of watermarks is fixed in advance, some work was also done on dynamic (adaptive) schemes. Besides the well-known deterministic scheme of Fiat and Tassa~\cite{fiat01} which requires a large bandwidth, Tassa~\cite{tassa05} constructed a low bandwidth dynamic scheme with a length of $O(K^4 \log N)$. Recently, Laarhoven et al.~\cite{laarhoven13tit} gave a more efficient dynamic scheme where the number of segments is only $O(K^2 \log N)$, and in~\cite{laarhoven12wifs} a trade-off construction was given to build schemes that require a higher bandwidth but with a smaller constant $T$. In these schemes, all colluders are caught with high probability, whereas in non-adaptive schemes, at least one colluder is caught with high probability.

\paragraph{Relation to group testing}
Oosterwijk et al.~\cite{oosterwijk13} recently considered optimizing Tardos' scheme to the scenario where the pirate strategy is known, e.g., when the pirates always randomly choose one of their versions (the \textit{interleaving} attack) or when the pirates always output the same watermarked version if at least one of them received this version (the \textit{all-$1$} attack). The latter pirate strategy corresponds to getting a pirate output of $1$ if and only if at least one traitor is present in the set of users who received a $1$. This shows that the traitor tracing game with the all-$1$ attack is in fact equivalent to the group testing game, and more generally that many group testing models correspond to specific pirate strategies in the traitor tracing game. Traitor tracing can therefore be seen as a generalization of group testing, or group testing as a special case of traitor tracing.

%..............................................................................%

\subsection{Contributions}

In this paper, we will show that combining and improving several of the aforementioned results from traitor tracing~\cite{tardos03, oosterwijk13, laarhoven13tit, laarhoven13wifs} leads to a group testing framework that can deal with many different group testing models efficiently. The resulting group testing algorithms we present are computationally efficient and, for sufficiently large $K$, require fewer tests than many known algorithms from the literature. For large $N$, the number of tests required in our schemes scales as follows, depending on the model. Here $r$ is a noise-parameter, which informally corresponds to the probability of not getting the expected result.
\begin{itemize}
  \item Traditional group testing: $T \sim 2 K \ln N$.
  \item Noisy group testing (dilution): $T \sim 2 K \ln N / (1 - r)$.
  \item Noisy group testing (additive): $T \sim 2 K \ln N / (1 - \sqrt{2r})$.
  \item Threshold group testing (majority): $T \sim \pi K \ln N$.
  \item Threshold group testing (Bernoulli gap): $T \sim 4 K \ln N$.
  \item Threshold group testing (linear gap): $T \sim 2 K^2 \ln N$.
\end{itemize}
These asymptotics apply to both adaptive and non-adaptive group testing, but the first order terms are considerably smaller in adaptive group testing than in non-adaptive group testing. Although we have worked out the details for several models, this paper aims to provide a framework to efficiently deal with \textit{any} group testing model. For instance, for threshold group testing with small gaps we did not provide explicit formulas, but one may derive them as we will explain below.

Besides these improvements and this framework, one goal of this paper is to further stimulate a cooperation between the areas of group testing and traitor tracing, as these areas are surprisingly similar. Much work has been done in both areas in similar directions (combinatorial designs, probabilistic analyses, information-theoretic lower bounds), and although the connection between the two areas has been made a few times before (e.g.,~\cite{meerwald11b, %stinson00, 
colbourn10}), a further exchange of ideas may lead to improved results in both areas.

The outline of this paper is as follows. In Sect.~\ref{sec:framework} we provide the aforementioned framework to deal with arbitrary group testing models. Then, in Sect.~\ref{sec:ord}, \ref{sec:noise}, and~\ref{sec:threshold} we apply our results to some previously considered models and present our results. Finally, in Sect.~\ref{sec:conc} we conclude by mentioning an important open problem in traitor tracing that might be of interest to the group testing community. All proofs and many details are omitted due to space limitations, but will appear in the full version.

%++++++++++++++++++++++++++++++++++++++++++++++++++++++++++++++++++++++++++++++%

\section{SCORE-BASED GROUP TESTING}
\label{sec:framework}

In this section, we will look at a framework for probabilistic group testing with average-case errors. We will cover both adaptive and non-adaptive group testing. Before introducing this framework, we first introduce some more notation. We write $X$ to denote the group testing matrix, or code matrix, indicating which items are included in which tests. We denote its length by $T$, which we will also call the code length. We denote test outcomes with $y$. Throughout, we will generally index items with $j$ and tests with $i$, i.e., $y_i$ is the outcome of the $i$th test, and $X_{j,i} = 1$ if and only if item $j$ is included in the $i$th test. Finally, we write $\eps_1$ for an upper bound on the probability that one or more non-defective items are marked as defective by our algorithm (getting one or more \textit{false positives}), and $\eps_2$ for an upper bound on the probability that some defective item is not marked defective (a \textit{false negative}).

%..............................................................................%

\subsection{Non-adaptive group testing}

In 2003, Tardos~\cite{tardos03} introduced a collusion-resistant traitor tracing scheme, which he showed to be order-optimal in the number of segments needed. In group testing terminology, this scheme relies on assigning test scores to items based on the results of each test, such that if we add up all test scores for each item, defective items will eventually get much higher scores than non-defective items. Given a certain probability $p$, a score function $h$, and a threshold $Z$, this scheme works as described in Fig.~\ref{fig:non}. Here $i$ refers to the $i$th test, and $j$ refers to the $j$th item. \footnote{In traitor tracing, one actually has to use varying values of $p$ for different tests, as otherwise an adversary could gain knowledge about $p$ and use this knowledge to build a strong attack that can beat the system. However, in group testing there is no adversary, and therefore there is no reason to vary $p$ for different tests.}
  
\begin{figure}
\begin{mdframed}[linewidth=1pt]
\noindent\textbf{Constructing the group test matrix $X$:}
  \begin{itemize}
    \item For each $i,j$, set $X_{j,i} = 1$ with probability $p$.
  \end{itemize}

\noindent\textbf{Finding the defectives, given the test results $y$:}
  \begin{itemize}
    \item For each $i,j$, calculate a \textit{score} $S_{j,i} = h(X_{j,i}, y_i)$.
    \item For each $j$, compute the total score $S_j = \sum_i S_{j,i}$. 
    \item Mark item $j$ as defective if and only if $S_j > Z$.
  \end{itemize}
\end{mdframed}
\caption{The general outline of non-adaptive score-based group testing. The parameters $p$, $h$, $Z$, and $T$ depend on the model and will be discussed later. \label{fig:non}}
\end{figure}

For the time being we develop the theory for a generic score function $h$, but it is generally chosen such that it assigns positive scores to matches ($X_{j,i} = y_i$) and negative scores to differences, and gives large positive (negative) scores to the matches (differences) that were the least likely. For each test, the expected score for a non-defective item is usually $0$, while for defective items it is strictly positive. Therefore, by running sufficiently many tests, with high probability we are able to distinguish between the scores of non-defective items (which have mean $0$) and the scores of defective items (which have a large positive mean).

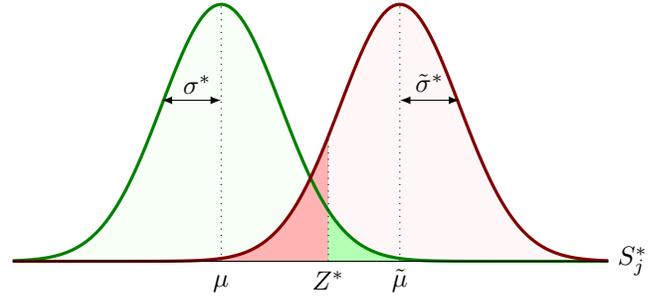
\begin{figure}
\begin{tikzpicture}
\begin{axis}[
  no markers, domain=0:10, samples=200, hide y axis,
  axis lines*=left, xlabel=$S_j^{*}$,
  every axis y label/.style={at=(current axis.above origin),anchor=south},  
  every axis x label/.style={at=(current axis.right of origin),anchor=west},
  height=5cm, width=1.1\columnwidth,
  xtick=\empty, ytick=\empty,
  enlargelimits=false, clip=false, axis on top,
  grid = major
  ]

%----------------------------
% Drawing non-defective curve
\addplot [fill=red!3, draw=none, domain=5.3:10] {gauss(6.5,1)} \closedcycle;
\addplot [fill=green!3, draw=none, domain=0:5.3] {gauss(3.5,1)} \closedcycle;
\addplot [fill=red!30, draw=none, domain=0:5.3] {gauss(6.5,1)} \closedcycle;
\addplot [fill=green!30, draw=none, domain=5.3:10] {gauss(3.5,1)} \closedcycle;

\addplot [very thick,green!50!black] {gauss(3.5,1)};

\draw [dotted] (axis cs:3.5,0) -- (axis cs:3.5,0.4);
\node at (axis cs:3.5,-0.035) {$\mu$};

\draw [latex-latex] (axis cs:2.5,0.25) -- (axis cs:3.5,0.25);
\node at (axis cs:3.1,0.275) {$\sigma^{*}$};

%------------------------
% Drawing defective curve

\addplot [very thick,red!50!black] {gauss(6.5,1)};

\draw [dotted] (axis cs:6.5,0) -- (axis cs:6.5,0.4);
\node at (axis cs:6.5,-0.03) {$\tilde{\mu}$};

\draw [latex-latex] (axis cs:6.5,0.25) -- (axis cs:7.5,0.25);
\node at (axis cs:7,0.277) {$\tilde{\sigma}^{*}$};

%------------------------
% Drawing the threshold Z
\draw [dotted] (axis cs:5.3,0) -- (axis cs:5.3,0.2);
\node at (axis cs:5.3,-0.03) {$Z^{*}$};

\end{axis}

\end{tikzpicture}

\caption{The Gaussian approximation of the score curves $S_j^{*} = S_j/T$, for non-defectives (left) and defectives (right). The means $\mu$ and $\tilde{\mu}$ do not depend on $T$, but $\sigma^{*} = \frac{\sigma}{\sqrt{T}}$ and $\tilde{\sigma}^{*} = \frac{\tilde{\sigma}}{\sqrt{T}}$ decrease when $T$ increases. For sufficiently large $T$, choosing $Z^{*}$ appropriately between $\mu$ and $\tilde{\mu}$ guarantees that the left (right) marked area has size at most $\frac{\eps_2}{K}$ ($\frac{\eps_1}{N}$). \label{fig:gauss}}
\end{figure}

To analyze the performance of score-based schemes, we need to estimate the probabilities that (a) a non-defective item is still marked as defective, and (b) a defective item is not marked as defective. To do this, first note that for each item $j$, the scores for each test $i$ are independently and identically distributed. For convenience, let us introduce the following notations for the mean and variance of the scores of non-defective items and defective items for each test. Below, we omit subscripts $i$ on $y$, and we use $x$ ($\tilde{x}$) to denote the symbol $X_{j,i}$ for non-defectives (defectives). Throughout, we will consistently use tildes to indicate variables corresponding to defective items.
\begin{align}
\mu &= \expn[h(x,y)], & \tilde{\mu} &= \expn[h(\tilde{x},y)], \\
\sigma^2 &= \var[h(x,y)], & \tilde{\sigma}^2 &= \var[h(\tilde{x},y)].
\end{align}
Now, the total score of an item $j$ is given by $S_j = \sum_i S_{j,i}$, where $S_{j,i} = h(X_{j,i}, y_i)$. This is a sum of many i.i.d. random variables, and due to the Central Limit Theorem, for large $T$ we expect $S_j$ to be approximately normally distributed with mean $\mu T$ ($\tilde{\mu} T$) and variance $\sigma^2 T$ ($\tilde{\sigma}^2 T$). So if we look at the average score per test $S_j^{*} = \frac{S_j}{T}$, non-defective items (defective items) will have a mean of $\mu$ ($\tilde{\mu}$) and a standard deviation of $\sigma^{*} = \frac{\sigma}{\sqrt{T}}$ ($\tilde{\sigma}^{*} = \frac{\tilde{\sigma}}{\sqrt{T}}$), as shown in Fig.~\ref{fig:gauss}. Therefore, when $\mu < \tilde{\mu}$ and $\sigma$ and $\tilde{\sigma}$ are sufficiently small, increasing $T$ will make both curves more narrow, and allow us to distinguish between the two curves with high probability. Working out the details, this leads to the following result about $T$ and $Z$. The proof, as well as many other details, can be found in the full version of this paper.

\begin{theorem} \label{thm:non-gauss}
Suppose we use the score-based non-adaptive group testing scheme described in Fig.~\ref{fig:non}, and the average item scores for each item follow a perfect Gaussian curve. Then, to guarantee that (i) a non-defective item is marked defective with probability at most $\frac{\eps_1}{N}$, and (ii) a defective is marked as non-defective with probability at most $\frac{\eps_2}{K}$, the following parameters suffice:
\begin{align}
T &= \frac{2}{(\tilde{\mu} - \mu)^2} \left[\sigma \sqrt{\logn} + \tilde{\sigma} \sqrt{\logk}\right]^2,
\end{align}
\begin{align}
%Z &= Z^{*} \cdot T = \frac{2\sigma^2}{(\tilde{\mu} - \mu)^2} \logn \left[1 + \frac{\tilde{\sigma}}{\sigma} \eta \right] \left[\tilde{\mu} + \frac{\tilde{\sigma} \mu}{\sigma} \eta \right] \\
Z &= \left(\frac{\sigma \tilde{\mu} \sqrt{\logn} + \tilde{\sigma} \mu \sqrt{\logk}}{\sigma \sqrt{\logn} + \tilde{\sigma} \sqrt{\logk}}\right) \cdot T.
\end{align}
In particular, it then follows that with probability at least $1 - \eps_1$, all non-defectives are correctly classified as non-defective, and with probability at least $1 - \eps_2$, all defective items are correctly marked as defective.
\end{theorem}

For notational convenience, let us write
\begin{align}
A = \frac{2\sigma^2}{(\tilde{\mu} - \mu)^2}, \quad B = \frac{\tilde{\sigma}}{\sigma}, \quad \etan = \sqrt{\frac{\ln(K/\eps_2)}{\ln(N/\eps_1)}},
\end{align}
so that the formula for the parameter $T$ in Thm.~\ref{thm:non-gauss} can be concisely expressed as
\begin{align}
T = A \logn \left[1 + B \etan\right]^2.
\end{align}
We generally have $\etan \leq 1$, while for small $K$ and large $N$, the value of $\eta$ will be very small. In fact, for $K = N^{o(1)}$ and $N \to \infty$ we have $\etan = o(1)$, leading to the following corollary.

\begin{corollary} \label{cor:non}
Suppose that $K = N^{o(1)}$, that $\eps_1$ and $\eps_2$ are fixed, and that $B = O(1)$. Then, for large $N$ we have
\begin{align}
T &\sim A \ln N, \qquad Z \sim A \tilde{\mu} \ln N.
\end{align}
\end{corollary}

To minimize the number of tests, we are therefore mostly aiming to minimize the value of $A$. This parameter depends on the choices of $p$ and $h$, and on the model of how the test result $y$ is produced. 

%..............................................................................%

\subsection{Adaptive group testing}

The procedure described in Fig.~\ref{fig:non} can be adapted to the adaptive setting by making the following small modification: instead of only marking items defective if their \textit{final} scores exceed $Z$, we mark an item defective (and do not include it in any of the remaining group tests) as soon as its score exceeds the threshold $Z$. This modification was recently proposed in~\cite{laarhoven13tit} to build efficient adaptive traitor tracing schemes from Tardos' non-adaptive scheme, but can also be used to make score-based group testing work even more efficiently. The modified scheme is presented in Fig.~\ref{fig:ada}. 

\begin{figure}
\begin{mdframed}[linewidth=1pt]
\noindent\textbf{Constructing $X$ and finding the defectives:} \\
For each $i = 1, \dots, T$, sequentially do the following. (Initially $\act = \{1, \ldots, N\}$ and $S_j(0) = 0$ for all $j$.)
  \begin{itemize}
    \item For each $j \in \act$, set $X_{j,i} = 1$ with probability $p$.
    \item Run the test, and obtain the test output $y_i$.
    \item For each $j \in \act$, do the following:
    \begin{itemize}
      \item Compute $S_{j,i} = h(X_{j,i}, y_i)$.
      \item Update $S_j(i) = S_j(i-1) + S_{j,i}$.
      \item Mark $j$ as defective if $S_j(i) > Z$.
    \end{itemize}
    \item Remove all items marked defective from $\act$.
  \end{itemize}
\end{mdframed}
\caption{\footnotesize How to adapt the non-adaptive score-based group testing schemes to the adaptive setting, and gain the factor $\sqrt{K}$ in the first order error term.\label{fig:ada}}
\end{figure}

It was shown in~\cite{laarhoven13tit,laarhoven13wifs} that with this modification, proving that the \textit{average} defective item score exceeds $Z$ is roughly enough to prove that \textit{all} defective items are found. This means that instead of looking at scores of \textit{single} defective items, we should now look at the \textit{average} score of all defective items. Compared to the right curve in Fig.~\ref{fig:gauss}, this curve has the same mean $\tilde{\mu}$, but because it is an averaged score over $K$ individual scores, the normalized standard deviation $\sigma^*$ will be $\sqrt{K}$ times smaller. This leads to the following result, a proof of which can be found in the full version.

\begin{theorem} \label{thm:ada-gauss}
Suppose that we use the score-based adaptive group testing scheme described in Fig.~\ref{fig:ada}, and suppose that the average item scores of all items follow a perfect Gaussian curve. Then, to guarantee that (i) a non-defective item is marked defective with probability at most $\eps_1/N$, and (ii) a defective item is not marked defective with probability at most $\eps_2/K$, the following parameters suffice:
\begin{align}
T &= \frac{2}{(\tilde{\mu} - \mu)^2} \left[\sigma \sqrt{\logn} + \frac{\tilde{\sigma}}{\sqrt{K}} \sqrt{\logk}\right]^2 \\
%Z &= Z^{*} \cdot T = \frac{2\sigma^2}{(\tilde{\mu} - \mu)^2} \logn \left[1 + \frac{\tilde{\sigma}}{\sigma} \eta \right] \left[\tilde{\mu} + \frac{\tilde{\sigma} \mu}{\sigma} \eta \right] \\
Z &= \left(\frac{\sigma \tilde{\mu} \sqrt{\logn} + \frac{\tilde{\sigma}}{\sqrt{K}} \mu \sqrt{\logk}}{\sigma \sqrt{\logn} + \frac{\tilde{\sigma}}{\sqrt{K}} \sqrt{\logk}}\right) \cdot T.
\end{align}
\end{theorem}

Similar to the non-adaptive group testing setting, we now write $\etaa = \sqrt{\frac{\ln(K/\eps_2)}{K \ln(N/\eps_1)}}$ so that the formula for the parameter $T$ in Thm.~\ref{thm:ada-gauss} can be concisely expressed as
\begin{align}
T = A \logn \left[1 + B \etaa\right]^2.
\end{align}
The parameter $\etaa$ is generally really small due to the factor $\sqrt{K}$. So without making any assumptions on $K$ and $N$, we may already claim that for large $K$ and/or $N$, the parameter $\etaa$ will go to $0$.

\begin{corollary} \label{cor:ada}
Suppose that $\eps_1$ and $\eps_2$ are fixed, and that $B = O(1)$. Then, for large $N$, we have
\begin{align}
T &\sim A \ln N, \qquad Z \sim A \tilde{\mu} \ln N
\end{align}
\end{corollary}

To summarize, the asymptotics of $T$ will generally be the same as in the non-adaptive model, but the convergence to this limit will be much faster due to the extra factor $\sqrt{K}$. Also, as noted in~\cite{laarhoven13tit}, the actual number of tests needed to find all defectives is generally much less than the theoretical upper bounds suggest.

%..............................................................................%

\subsubsection{Dealing with unknown $K$}

In \cite{laarhoven13tit}, a scheme is discussed to effectively deal with adaptive scenarios where the number of defectives is not known in advance (the \textit{universal} Tardos scheme), while maintaining equivalent asymptotics on $T$. This roughly comes down to using several thresholds $Z$, and the same idea may also be applied to adaptive group testing with an unknown number of defectives. For details, see~\cite[Sect.~V]{laarhoven13tit}.

%..............................................................................%

\subsubsection{Reducing the number of stages}

In \cite{laarhoven13tit}, a setting somewhere between non-adaptive and adaptive traitor tracing is also discussed (the \textit{weakly dynamic} Tardos scheme), and how one could adapt the adaptive scheme to such a setting effectively. Translating those results to group testing, the same asymptotics on $T$ hold even if the number of stages is reduced to $O(K)$ (with $O(T/K)$ tests in each stage). But reducing the number of rounds does lead to larger first order terms and larger practical code lengths. For details, see~\cite[Sect.~IV]{laarhoven13tit}.

%..............................................................................%

\subsection{Optimal score functions $h$}

Recently, Oosterwijk et al.~\cite{oosterwijk13} studied the score functions used in traitor tracing, and showed that if the attack strategy of the pirates is known, then the score function $h$ that minimizes $A$ is given as follows. This choice of $h$ is such that it is both \textit{centered} ($\mu = 0$) and \textit{quasi-normalized} ($\sigma^2 = \tilde{\mu}$).

\begin{lemma} \cite[Cor.~6]{oosterwijk13}
The optimal, centered ($\mu = 0$) and quasi-normalized ($\sigma^2 = \tilde{\mu}$) score function $h$ that minimizes $A$ under the Gaussian assumption is given by
\begin{align}
h(x,y) = \frac{1}{K} \left.\frac{\partial \ln\left(p_{y \mid p_0, p_1}\right)}{\partial p_x}\right|_{p_1 = 1 - p_0 = p} \, ,
\end{align}
where $p_{y} = P(y_i = y)$ and $p_x = P(X_{j,i} = x)$.
\end{lemma}

For several attack strategies explicit formulas for $h$ were derived in~\cite{oosterwijk13}, some of which we will encounter later. For one particular strategy they obtained a score function that turned out to achieve capacity in the non-adaptive traitor tracing game.

%..............................................................................%

\subsection{Optimal probabilities $p$}

Once the model (in traitor tracing: attack) is fixed, we can now compute the optimal score function $h$ as described above, and we are almost done. To finalize the scheme, we then only need to choose a parameter $p$. Since the parameters $A$ and $B$, and therefore a Gaussian-based estimate of the code length $T$, can be explicitly computed as a function of $p$, what remains is a straightforward optimization of $p$ minimizing the estimate of $T$. Asymptotically, as shown in Cor.~\ref{cor:non} and \ref{cor:ada}, we would like to choose $p$ so as to minimize $A$, but in practice there is a trade-off between minimizing $A$ and minimizing $B$. We will further discuss this below.

%++++++++++++++++++++++++++++++++++++++++++++++++++++++++++++++++++++++++++++++%

\section{TRADITIONAL GROUP TESTING}
\label{sec:ord}

With the framework in place, we are ready to start building group testing schemes in arbitrary models, and we will discuss the results in the next few sections. We will naturally start with the most often considered, traditional group testing model, where the outcome of a test is positive if and only if at least one defective item is present in the tested pool. We will first give a scheme based on a straightforward optimization of $h$ and $p$, and then discuss how the score function can be slightly refined in this particular model, leading to smaller constants $T$. 

%..............................................................................%

\subsection{The direct approach}

First, Oosterwijk et al.~\cite[Cor.~22]{oosterwijk13} showed that the following centered and quasi-normalized score function is optimal in the ordinary group testing model, in that it minimizes $A$ under the Gaussian assumption. 
\begin{align}
%h(x,y) &= (-1)^{x+y} \left(\frac{p}{1 - p}\right)^{1 - x} \left(\frac{(1 - p)^K}{1 - (1 - p)^K}\right)^{y}.
h(x,y) &= \begin{cases} 
+p/(1 - p) 							& (x,y) = (0,0) \\
-p(1 - p)^{K - 1}/(1 - (1 - p)^K) \hspace{-0.2cm} & (x,y) = (0,1) \\
-1 									& (x,y) = (1,0) \\ 
+(1 - p)^K/(1 - (1 - p)^K)			& (x,y) = (1,1) 
\end{cases}
\end{align}
Using this score function, we can compute the parameters $A$ and $B$ as a function of $p$, and find the optimal value of $p$ that minimizes $T$. For arbitrary values of $K$, these parameters are somewhat ugly functions of $p$, but the optimization of $p$ is just a straightforward procedure. These details can be found in the full version, but here we will focus on the cleaner asymptotics of $T$. Note that ``asymptotics" here refers to considering the case of large $K$, although the results may already provide good approximations of the actual value of $T$ when $K$ is small.

First, as is well known in group testing, one generally has to use small values of $p$ and sparse matrices $X$. It is generally assumed that $p = \frac{\alpha}{K}$ for some $\alpha$ which is constant or almost constant in $K$. Using the same parametrization here, we obtain the following asymptotics for the code length constants $A$ and $B$:
\begin{align}
A &= \frac{2(e^{\alpha} - 1)}{\alpha} K + O(1), \\ 
B &= \sqrt{\frac{1}{e^{\alpha} - 1}} + O\left(\frac{1}{K}\right).
\end{align}
Here, it should be noted that the leading term of $A$ is a strictly increasing function of $\alpha$, while the leading term of $B$ is strictly decreasing in $\alpha$. There is a clear trade-off between $A$ and $B$, and the optimal choice of $\alpha$ depends on the exact set of parameters $K$, $N$, $\eps_1$ and $\eps_2$. If we focus on the regime of large $K$, we see that $\alpha \to 0$ is optimal to minimize $A$, in which case we get
\begin{align}
T = 2 K \logn \left(1 + O\left(\alpha + \frac{\eta}{\sqrt{\alpha}}\right)\right).
\end{align}
Setting $\alpha = O(\eta^{2/3})$ balances the order terms, and leads to a first order term of the order $O(\eta^{2/3})$. But the important thing to note here is the leading term of $T$:
\begin{align}
T \sim 2 K \ln N.
\end{align}
For sufficiently large $K$, this improves upon results of Chan et al.~\cite{chan11}. It has to be noted that in their schemes, there are never any false positives (i.e., $\eps_1 = 0$), which is not true with the above construction. But if a small margin of error is present anyway (e.g., due to errors in the testing procedure), marked items may have to be tested individually anyway to confirm that the items are defective. In that case, allowing $\eps_1 > 0$ makes sense. Note that the asymptotics of $T$ are only a factor $2 \ln 2 < 1.39$ above the information-theoretic lower bound of Seb\H{o}~\cite[Thm.~2]{sebo85}. 

\begin{figure}[t]
	\centering
	\subfloat[][Non-adaptive traditional group testing]{\includegraphics[width=8.5cm]{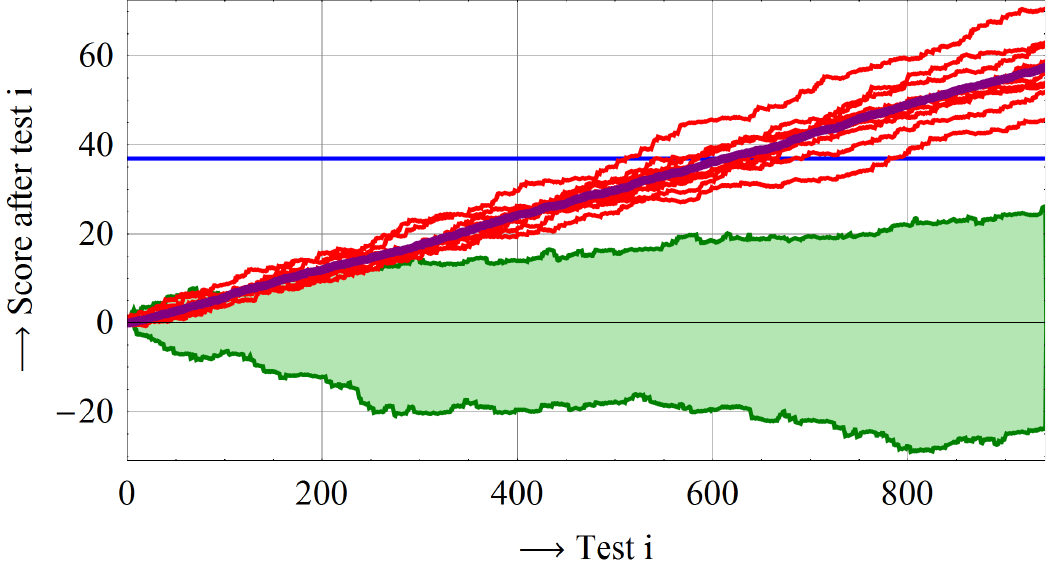} \label{fig:sim1a}} \\
	\subfloat[][Adaptive traditional group testing]{\includegraphics[width=8.5cm]{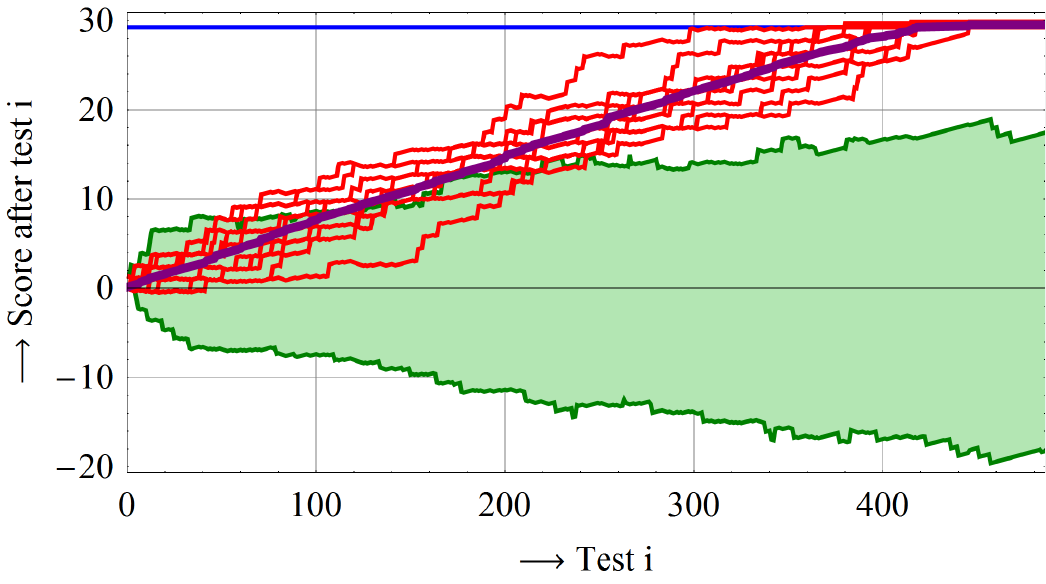} \label{fig:sim1b}}
	\caption{An example of the score-based scheme in action, in the non-adaptive setting and in the adaptive setting. The green marked area shows the range of all non-defective item scores, the red lines show the scores of defective items, the horizontal blue line shows the threshold $Z$, and the purple bold line shows the average score of defective items. As one can see, in non-adaptive group testing we need this purple line to really exceed $Z$, while in the adaptive setting it suffices to let this average hit $Z$. }
	\label{fig:sim1}
\end{figure}

To give an idea of how the scheme actually works, an example is given in Fig.~\ref{fig:sim1} with toy parameters $K = 10$, $N = 1000$, and $\eps_{1,2} = 10^{-2}$. For the non-adaptive scheme, optimizing $p$ then gives us $p \approx 0.091$ leading to $T \approx 941$ and $Z \approx 37$, while for the adaptive setting we get $p \approx 0.055$ with $T \approx 486$ and $Z \approx 29$. \footnote{Note that in noiseless group testing, a trivial binary search leads to a much better adaptive scheme with $\eps_{1,2} = 0$ and $T = K \lceil \log_2 N \rceil = 100$ tests. However, in noisy settings (Sect.~\ref{sec:noise}) and several other models (Sect.~\ref{sec:threshold}) such trivial algorithms do not exist, and in those cases our adaptive construction may also be of interest.}

%..............................................................................%

\subsection{Fine-Tuning the Score Function}
\label{sub:finetune}

Taking a step back from the score-based construction and looking at the traditional group testing model, we know that if an item is included in a test ($x = 1$) while the test result is negative ($y = 0$), this item is \textit{certainly} not defective. So in those cases, instead of assigning this item a somewhat negative score of $-1$ (which may not be enough to guarantee that the item is not marked defective), we may also assign items a score of $-\infty$ when they are included in a test which comes back negative. So we may fine-tune $h$ by setting $h(1,0) = -\infty$. Then in each segment, with probability $q = p (1 - p)^K$ a non-defective item is assigned a score of $-\infty$. So with probability $1 - (1 - q)^T$ a non-defective has a score of $-\infty$ after $T$ segments. Setting $p \approx \frac{1}{K}$ maximizes the latter probability, as was previously noted in~\cite{chan11}, and eventually leads to an asymptotic code length of
\begin{align}
T \sim \frac{2 e (e - 1) K \ln N}{2e - 1} \approx 2.11 K \ln N.
\end{align}
So also for $p = \frac{1}{K}$ we end up with improved asymptotics for $T$, compared to the $T = e K \logn$ of~\cite{chan11}.

%++++++++++++++++++++++++++++++++++++++++++++++++++++++++++++++++++++++++++++++%

\section{NOISY GROUP TESTING}
\label{sec:noise}

We saw in Sect.~\ref{sub:finetune} that we may use the fact that the result of a test is never positive when one of the defective items is present in the pool, to fine-tune the score function and find all defectives even more efficiently. However, such certainties generally do not exist, as tests may have a small probability of not returning the correct result. Here we discuss two noisy group testing models previously considered in the literature, and show what the asymptotics on $T$ become. Details on how these results were obtained are in the full version.

%..............................................................................%

\subsection{Dilution model}

In the dilution model~\cite{atia09,atia12,cheraghchi09,cheraghchi11,sejdinovic10}, we assume that a test may not come back positive even if a defective item is present in the tested pool, because this defective item may be \textit{inactive} with a small probability $r$. This means that the probability that a defective item contributes a $1$ to the test result is now not $p$, but $p' = p(1 - r)$. In this model, optimizing $h$ leads to the centered and quasi-normalized score function given in Table~\ref{tab:1}. To minimize $A$, we again need to take $\alpha$ close to $0$, in which case the asymptotic code length becomes
\begin{align}
T \sim \frac{2 K \ln N}{1 - r} \, .
\end{align}
This is somewhat comparable to a result of~\cite{atia12} which has a factor $\frac{1}{(1 - r)^2} \approx \frac{1}{1 - 2r}$ in the denominator.

\begin{table*}[t]
\centering

    \caption{Optimal parameter choices for several group testing models, together with the resulting asymptotics on $T$. The plots on the left sketch $P(y = 1 \mid \beta)$ against $\beta$, where $\beta$ is the number of defectives included in a group test, for each of the given models. The different cases for $h$ always correspond to $(x,y) = (0,0), (0,1), (1,0)$, and $(1,1)$ respectively.}
    \label{tab:1}
\renewcommand{\arraystretch}{1.3}% Tighter
    \begin{small}
    \begin{tabular}{p{2.7cm}p{8.5cm}p{2.0cm}p{2.5cm}}
    \toprule 
    {\bfseries Model} & {\bfseries Optimal score function $h$} & {\bfseries Optimal $p$} & {\bfseries Asymptotics $T$} \\ 
    \midrule
	Traditional model & \multirow{4}{*}{$h(x,y) = \begin{cases} 
	+p/(1 - p) 							 \\
	-p(1 - p)^{K - 1}/(1 - (1 - p)^K) 	\\
	-\infty 							\\ 
	+(1 - p)^K/(1 - (1 - p)^K)			
	\end{cases}$} & \multirow{2}{*}{$p = \dfrac{o(1)}{K}$} & \multirow{2}{*}{$T \sim 2 K \ln N$} \\
	\multirow{3}{*}{\includegraphics[width=2.5cm]{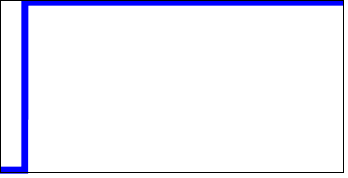}}			& & & \\ %\cline{3-4}
			& & \multirow{2}{*}{$\Big(p = \dfrac{1}{K}\, ,$} & \multirow{2}{*}{$T \sim 2.11 K \ln N\Big)$} \\
			& & & \\
    \midrule
    Noise: Dilution & \multirow{4}{*}{$h(x,y) = \begin{cases} 
+p(1 - r)/(1 - p(1 - r)) 							\\
\text{\scriptsize $-p(1 - r)(1 - p(1 - r))^{K - 1}/(1 - (1 - p(1 - r))^K)$} 	\\
-1 + r/(1 - p(1 - r))						\\ 
\text{\scriptsize $+(1 - p(1 - r))^{K-1}(1 - p)(1 - r)/(1 - (1 - p(1 - r))^K)$}
\end{cases}$} &	\multirow{4}{*}{$p = \dfrac{o(1)}{K}$} & \multirow{4}{*}{$T \sim \dfrac{2 K \ln N}{1 - r}$} \\
	\multirow{3}{*}{\includegraphics[width=2.5cm]{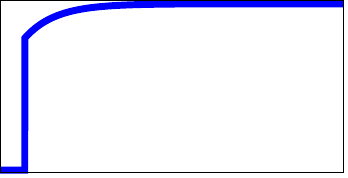}}		& & & \\
			& & & \\ 
			& & & \\
    \midrule
    Noise: Additive & \multirow{4}{*}{$h(x,y) = \begin{cases} 
+p/(1 - p) \\
-p (1 - p)^{K - 1} (1 - r)/(1 - (1 - p)^K (1 - r)) \\
-\infty \\ 
+(1 - p)^K (1 - r)/(1 - (1 - p)^K (1 - r))
\end{cases}$} &	\multirow{4}{*}{$p \approx \dfrac{\sqrt{2r}}{K}$} & \multirow{4}{*}{$T \sim \dfrac{2 K \ln N}{1 - \sqrt{2r}}$} \\
	\multirow{3}{*}{\includegraphics[width=2.5cm]{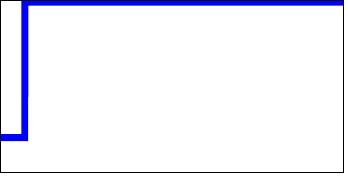}}		& & & \\
			& & & \\ 
			& & & \\
    \midrule
    Threshold: Majority & \multirow{4}{*}{$h(x,y) = \begin{cases} 
	+1  \\
	-1 	\\
	-1 	\\ 
	+1	
	\end{cases}$} &	\multirow{4}{*}{$p = \dfrac{1}{2}$} & \multirow{4}{*}{$T \sim \pi K \ln N$} \\
	\multirow{3}{*}{\includegraphics[width=2.5cm]{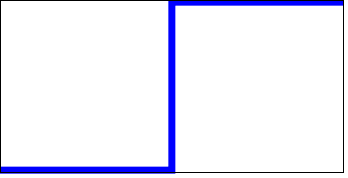}}		& & & \\
			& & & \\
			& & & \\
    \midrule
    Threshold: Bernoulli & $h(x,y) = \left(p^{K-1} + (1 - p)^{K-1}\right)$ & \multirow{4}{*}{$p = \dfrac{o(1)}{K}$} & \multirow{4}{*}{$T \sim 4 K \ln N$} \\
	\multirow{4}{*}{\includegraphics[width=2.5cm]{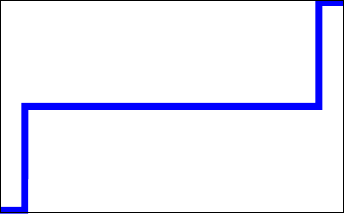}}		& \multirow{4}{*}{$\qquad \quad \ \times \begin{cases}
	+p/(1 - p^K + (1 - p)^K) \\ 
	-p/(1 + p^K - (1 - p)^K)				\\ 
	-(1 - p)/(1 - p^K + (1 - p)^K)			\\ 
	+(1 - p)/(1 + p^K - (1 - p)^K)			\\ 
	\end{cases}$} & & \\ 
   			& & & \\ 
			& & & \\
			& & & \\
    \midrule
    Threshold: Linear & \multirow{4}{*}{$h(x,y) = \begin{cases} 
	+p/(1 - p) \\
	-1 			\\
	-1 			\\ 
	+(1 - p)/p	 
	\end{cases}$} & \multirow{4}{*}{$p = \dfrac{1}{2}$} & \multirow{4}{*}{$T \sim 2 K^2 \ln N$} \\
    \multirow{3}{*}{\includegraphics[width=2.5cm]{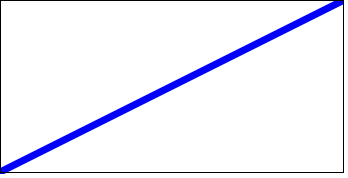}}	& &	& \\ %\cline{3-4}
            & &	& \\
            & & & \\
    \bottomrule
    \end{tabular}
    \end{small}
\end{table*}

%..............................................................................%

\subsection{Additive model}

Another commonly considered noise model is that of additive noise~\cite{atia09,atia12,sejdinovic10}, where the final extraction of the test result may not always be correct. In particular, we assume that we are in the ordinary group testing model, but the output $y$ may also be $1$ with probability $r$ if no defectives were actually present in the test. For this model, after fine-tuning $h$ we get the score function given in Table~\ref{tab:1}. To optimize the leading term of $A$ for fixed $r > 0$, one should choose $\alpha$ to satisfy $e^{\alpha} (1 - \alpha) = 1 - r$, which corresponds to $\alpha \approx \sqrt{2r} + O(r)$. For the asymptotics of the code length we then get
\begin{align}
T \sim \frac{2 K \ln N}{1 - \sqrt{2r} + O(r)}\, .
\end{align}
Note that Atia and Saligrama~\cite{atia12} showed that a code length of the order $O(\frac{K \log N}{1 - r})$ is already sufficient, and that our result, although practical, does not achieve this bound.

%++++++++++++++++++++++++++++++++++++++++++++++++++++++++++++++++++++++++++++++%

\section{THRESHOLD GROUP TESTING}
\label{sec:threshold}

Finally, a model that has also been considered before is \textit{threshold} group testing~\cite{damaschke06,chan13}, where a test result may only be positive if sufficiently many defective items are present in the tested pool. We will restrict our attention to the case where the test result is a (non-deterministic) function only of the number of defectives in a tested pool. This means that all defectives are treated symmetrically, and the test result does not depend on how many non-defectives were present in the tested group. In all models, it is assumed that: if at most $\ell$ defectives are present in a test, the output will be negative; if at least $u$ defectives are present, the test result is positive; and if the number of defectives $\beta$ in a group test lies between $\ell + 1$ and $u - 1$, the result depends on the specific model. Details on the following results can be found in the full version.

%..............................................................................%

\subsection{Majority group testing}

This model was introduced in~\cite{lebedev10}, and considers the case where $y = 1$ if and only if more than half the defectives are present in the tested pool. This corresponds to $\ell = \frac{K - 1}{2}$ and $u = \frac{K + 1}{2}$. In this case, the score function $h$ becomes a mess, but not if we immediately set $p$ to its optimal value, which turns out to be $p = \frac{1}{2}$. In that case, the score function reduces to the trivial function of $+1$ for matches and $-1$ for differences, as shown in Table~\ref{tab:1}. Working out the details for large $K$ leads to an asymptotic code length of
\begin{align}
T \sim \pi K \ln N.
\end{align}
Interpolating between the ordinary group testing model and majority group testing, one might expect that if $\ell = u - 1$ with $0 < \ell < \frac{K - 1}{2}$, the optimal value of $p$ is around $\frac{\ell}{K}$ and the asymptotics on $T$ are between $2 K \ln N$ and $\pi K \ln N$. 

%..............................................................................%

\subsection{Bernoulli model}

The Bernoulli gap model was previously considered in~\cite{chan13}, and says that if the number of defectives in a pool is between $\ell + 1$ and $u - 1$, the probability that the test outcome is positive equals $q = \frac{1}{2}$. We will focus on the extreme case of $\ell = 0$ and $u = K$, although a similar analysis may be done for other values of $\ell$ and $u$. First, the optimal score function follows from~\cite[Cor.~22]{oosterwijk13} and is given in Table~\ref{tab:1}. As in ordinary group testing, the optimal value of $p$ lies close to $0$, and for large $K$ the asymptotic scaling of $T$ is given by
\begin{align}
T \sim 4 K \ln N.
\end{align}
Again, interpolating between several results, if the gap between $\ell$ and $u$ decreases, we conjecture that the constant $T$ goes down from $4 K \ln N$ to $2 K \ln N$ if $\ell \to u = K$ or $u \to \ell = 0$, and from $4 K \ln N$ to $\pi K \ln N$ if $\ell, u \to \frac{K}{2}$.

%..............................................................................%

\subsection{Linear model}

In the linear gap model~\cite{chan13,dellungo05}, the probability of the test result to be positive scales linearly with the number of defectives in the tested pool. We will again only focus on the case of an extreme gap ($\ell = 0$ and $u = K$) for ease of computation. First, the optimal centered and normalized score function follows from~\cite[Prop.~9]{oosterwijk13} and is given in Table~\ref{tab:1}. As shown in \cite[Prop.~10]{oosterwijk13}, for this model we have $\tilde{\mu} = \frac{1}{K}$ regardless of $p$, so the best we can do is choose $p$ such that $\tilde{\sigma}^2$ is minimized. This leads to $p = \frac{1}{2}$ and $\tilde{\sigma}^2 = 1 - \frac{1}{K^2}$, and the asymptotic code length becomes
\begin{align}
T \sim 2 K^2 \ln N.
\end{align}
For large $N$ this slightly improves upon a previous result of Del Lungo et al.~\cite{dellungo05}, who gave an adaptive scheme with a code length of $T \sim 2 K^2 \log_2 N > 2.88 K^2 \ln N$.

%..............................................................................%

\subsection{Unknown model}

Finally, if we assume that the output will be a $0$ if no defectives are present, the output is $1$ if all defectives are present, and we do not know what happens when \textit{some} defectives are included in the test, then we are back at the traitor tracing game. For this game it is known that in the non-adaptive setting, the capacity-achieving choice is to use the same score function as in the linear gap model, but to vary $p$ for each test by independently drawing it each time from the arcsine distribution (with distribution function $F(p) = \frac{2}{\pi} \arcsin \sqrt{p}$ on $(0,1)$). This leads to the so-called interleaving defense, discussed in~\cite{oosterwijk13, laarhoven13wifs}, with an asymptotic code length of $T \sim 2 K^2 \ln N$. This result is the same as in the linear gap model, which motivates why the linear gap model is the hardest group testing model to deal with. 

%++++++++++++++++++++++++++++++++++++++++++++++++++++++++++++++++++++++++++++++%

\section{CONCLUSION}
\label{sec:conc}

In this paper we considered a new framework for probabilistic non-adaptive and adaptive group testing schemes, based on combining several results from traitor tracing. This lead to efficient group testing schemes for various models.

Although in this work we applied results from traitor tracing to group testing, one may wonder whether something can be done in the other direction as well. With the recent traitor tracing result of~\cite{oosterwijk13} achieving capacity in the non-adaptive traitor tracing game, the latter game seems kind of ``solved". For the adaptive traitor tracing game one important open question remains, which is establishing the adaptive (dynamic) traitor tracing capacity. Not much is known about this yet, but perhaps combining previous techniques from adaptive group testing~\cite{aldridge12, baldassini13} and non-adaptive traitor tracing~\cite{huang12} may bring us closer to a solution.

%++++++++++++++++++++++++++++++++++++++++++++++++++++++++++++++++++++++++++++++%

\section*{ACKNOWLEDGMENT}

The author is grateful to Jan-Jaap Oosterwijk for many valuable suggestions and insightful discussions. The author further thanks Jeroen Doumen, Antonino Simone, Boris \v{S}kori\'{c}, and Benne de Weger for their valuable comments.

%++++++++++++++++++++++++++++++++++++++++++++++++++++++++++++++++++++++++++++++%

\end{document}